\documentclass[12pt]{article}

\usepackage{amssymb}
\usepackage{bm} 
 
\begin{document}
\begin{center}

{\LARGE  Spin dynamics with non--abelian Berry gauge fields 
as a semiclassical constrained hamiltonian system }

\vspace{17mm}

\"{O}mer F. Dayi\footnote{E-mail addresses: dayi@itu.edu.tr and
dayi@gursey.gov.tr.} 

\vspace{5mm}

\noindent{\it Physics Department, Faculty of Science and
Letters, Istanbul Technical University, TR-34469 Maslak--Istanbul, Turkey,} 

and

\noindent {\it Feza G\"{u}rsey Institute, P.O. Box 6,
TR--34684, \c{C}engelk\"{o}y--Istanbul, Turkey. } 

\end{center}

\vspace{1cm}

{\small
\noindent
The dynamics of observables which are
matrices depending on  $\hbar$ and taking values in classical phase space  
is defined  retaining the terms up to the first order in $\hbar$ of 
the Moyal bracket. 
Within this semiclassical approach a first order lagrangian
involving  gauge fields is studied as a
constrained hamiltonian system. This provides
a systematic study of spin dynamics 
in the presence of non--abelian Berry gauge fields. 
We applied the method to various types of dynamical
spin systems and clarified some persisting discussions. 
In particular employing the Berry gauge field which generates
the Thomas precession, we calculated the
force exerted on an electron in the external  electric and
magnetic fields. Moreover, a simple semiclassical formulation of the spin 
Hall effect is accomplished.}

\vspace{1cm}


\vspace{0.8cm}

\newcommand\be{\begin{equation}}
\newcommand\ee{\end{equation}}

\newcommand\no{\nonumber\\}
\newcommand\beqa{\begin{eqnarray}}
\newcommand\eeqa{\end{eqnarray}}
\newcommand\bea{\begin{eqnarray*}}
\newcommand\eea{\end{eqnarray*}}

\newcommand\tet{\theta}
\newcommand\la{\lambda}
\newcommand\lb{\label}
\newcommand\m{\mu}
\newcommand\n{\nu}
\newcommand\al{\alpha}
\newcommand\bet{\beta}
\newcommand\ga{\gamma}
\newcommand\de{\delta}
\newcommand\s{\sigma}
\newcommand\e{\epsilon}

\newcommand\ro{\rho}
\newcommand\ka{\kappa}
\newcommand\A{{\cal A}}
\newcommand\B{{\cal B}}
\newcommand\F{{\cal F}}
\newcommand\La{{\cal L}}
\newcommand\Ha{{\cal H}}
\newcommand\sy{{\cal S}}
\newcommand\cO{{\cal O}}

\newcommand\fr{\frac}
\newcommand\del{\partial}
\newcommand\h{\hat}

\newcommand\lp{\left(}
\newcommand\rp{\right)}


\newcommand\md{{\mathcal D}}

\newcommand{\ssc}{\stackrel{C}{\star}}
\newcommand\ti{\tilde}

\section{Introduction}

In  \cite{mnz} the
intrinsic spin Hall effect was studied considering
abelian and non--abelian Berry gauge fields\cite{berry}
arising from  the adiabatic transport.
After this seminal work
there has been a great effort to employ Berry gauge fields 
to acquire a better understanding of spin dependent dynamics semiclassically
\cite{bk}--\cite{ccn}.
Although similar phenomena were treated letting
coordinates and/or momenta be noncommuting, they appear to be disconnected.
We would like to present a formulation which embraces these approaches.
In our formulation
keeping track of the semiclassical approximation is easy and interactions between different
gauge fields can be introduced in a simple manner.

To present our approach we need to recall 
the Weyl--Wigner--Groenewold--Moyal (WWGM) method
of quantization\cite{wwgm}
 as well as the Dirac formulation of constrained hamiltonian
systems\cite{hrt}.

Quantum dynamics of particles without spin
is usually provided by operators depending on the
quantum phase space variables $(\h p_\mu , \h x_\mu)$ satisfying the Heisenberg algebra:
$[ \h p_\mu , \h x_\nu ]=-i\hbar\delta_{\mu\nu},$
$[\h p_\mu , \h p_\nu ]=0,$ $[\h x_\mu , \h x_\nu ]=0.$
However, there is an alternative
approach due to  WWGM where one introduces
symbols of operators and their star product:
  Observables are functions of
classical phase space  variables and operator product is replaced 
with star product\cite{wwgm}.
The WWGM method
works well for observables possessing a classical limit.
However, it is not clear
how it should be generalized to embrace  spin degrees of freedom.
Spin may  be incorporated into classical mechanics considering
the semiclassical approximation as well as the
nonrelativistic limit of the Dirac hamiltonian.
The latter is given in terms of operator valued matrices.
Hence, we consider observables which are matrices   whose
elements are functions of  classical phase space variables but depend on $\hbar$. 
Dynamical equations of these matrix--valued symbols
will be given by
a semiclassical bracket  acquired  from the Moyal bracket. 

When there are some different types of 
gauge fields, they can be incorporated into the hamiltonian formalism by
considering an enlarged  lagrangian system
which leads to second class constraints. 
Indeed,
to embed Berry gauge fields
in the semiclassical scheme a constrained hamiltonian dynamics 
will be presented starting from an appropriate matrix valued lagrangian.
We adopt the Dirac formulation of constrained hamiltonian dynamics 
replacing the Poisson bracket
with the proposed semiclassical bracket. 
This furnishes us with
a systematic formulation of dynamics when non--abelian Berry
gauge fields are present.

Once this formulation of matrix valued observables coupled to  gauge fields 
is accomplished we can employ it
to investigate dynamical properties of diverse spin systems.
  The semiclassical
dynamics of Bloch electrons in the  adiabatic approximation
where interband interactions are neglected was discussed
in \cite{xsn},\cite{b}.
We  study the same problem
within our approach. We achieved the correct phase space measure 
and noncommutativity of phase space variables.
Unitary transformations which generate Berry gauge fields are
also considered. We derived the equations of motion which can be
used in topological spin  transport\cite{bb}.  
In \cite{shen}  was shown that when an electric field is applied
to an electron,  a transverse force on the spin current occurs. 
This force resulted in the Heisenberg equation of motion
of  velocity, considering the nonrelativistic limit of the Dirac hamiltonian. 
We show that it can  easily be derived within our formulation. 
On the other hand the
nonrelativistic limit of the Dirac hamiltonian can also be obtained by a momentum dependent
Berry gauge field  which generates the 
Thomas precession\cite{mat}. We  calculate the force acting on an electron in
the electric
and magnetic fields in the presence of this gauge field.
The same transverse force on the spin current occurs which depends only
on the electric field. However, the terms depending both on the magnetic and electric fields  
do not concur. Experiments may   settle this disagreement.
The intrinsic
spin Hall effect was envisaged  in \cite{sin}
analyzing the spin current due to the Rashba hamiltonian\cite{ras}. 
Investigating the Rashba Hamiltonian 
is still attractive,
although when the vertex corrections are taken into account
the originally proposed universal behavior 
of spin Hall conductivity does not survive\cite{vsh}
(for a review see \cite{erh}). We study the  Rashba spin--orbit coupling
within our semiclassical approach 
to attain a very simple formulation of the spin Hall conductivity.
It is inspired by the derivation of the
Hall conductivity by demanding that the force
acting on electrons  vanishes.

In Section 2 we present the  semiclassical bracket of matrix valued
observables and its basic properties. The semiclassical constrained hamiltonian 
formulation which leads to a systematic approach of analyzing dynamical systems
with different sorts of non--abelian gauge fields is given in Section 3. An
application
of the formalism to
various  dynamical spin systems is  considered in Section 4. We clarified
some persisting discussions and also gave a simplistic formulation
of the spin Hall conductivity. The obtained results and other possible applications  are
discussed in the concluding section.

\section{Semiclassical symbols and the Moyal bracket}

Let us deal with the  classical 
canonical variables $(p_\mu , x_\mu )$ corresponding to the
quantum 
phase space   $(\h p_\mu , \h x_\mu);$
$\mu =1,\cdots ,M.$ In the  WWGM  method of quantization 
one considers the symbol map \cite{wwgm}
\be
\sy \left(\h f(\h p ,\h x ) \rp =f(  p , x ),
\ee
where  $f( p , x)$ is
the c--number 
function corresponding 
to the operator $\h f (\h p ,\h x ).$ 

Let the operator product of the quantum observables $\h f$ and $ \h g$ be 
$$
\h f(\h p ,\h x )\h g(\h p ,\h x )=\h h (\h p ,\h x ).
$$
Symbol map should respect the  operator product, so that 
we should introduce a (star) product satisfying
\be
\sy \lp \h f(\h p ,\h x )\h g (\h p ,\h x ) \rp =\sy \lp \h h (\h p ,\h x ) \rp =
\sy \Big( \h f(\h p ,\h x )\Big) \star
\sy \Big( \h g (\h p ,\h x ) \Big) .
\ee
Obviously the symbol map as well as the $*$--product
depend on the operator ordering adopted.
We deal with the Weyl ordering where the
associative star product is 
\be
\lb{star}
\star   =  
\exp \left[
\frac{i\hbar }{2} \left(
\frac{\overleftarrow{\del}}{\del x^\m}
\frac{\overrightarrow{\del}}{\del p_\m} 
-\frac{\overleftarrow{\del}}{\del p^\m}
\frac{\overrightarrow{\del}}{\del x_\m}  
\right) \right] . \label{sct}
\ee
The arrows on the derivatives indicate the direction in which they should be
applied. We adopt the Einstein convention, hence  the
repeated indices are summed over. To imitate the commutator of operators, we 
define the 
Moyal bracket of  two arbitrary observables $f(p,x )$ and $g(p,x )$ as
\be
\lb{oM}
[f(p,x ) , g(p,x ) ]_\star \equiv f(p,x ) \star g(p,x ) - g(p,x ) \star f(p,x ) .
\ee
Hence,
the classical phase space variables 
satisfy the Moyal bracket
\be 
[ p_\m, x^\n ]_\star = -i\hbar \delta^\n_\m ,
\ee
analogous to the canonical commutation relations.
The classical limit  of the Moyal bracket (\ref{oM}) is
the Poisson bracket:
\be
{\lim_{\hbar \rightarrow 0}} \fr{-i}{\hbar } [f(p,x ) , g(p,x ) ]_\star 
= \{ f(p,x ) , g(p,x ) \} \equiv
  \frac{\partial f}{\partial x^\n} \frac{\partial g}{\partial p_\n}
-\frac{\partial f}{\partial p_\n}\frac{\partial g}{\partial x^\n} . \lb{ocl}
\ee

When one considers the  Dirac hamiltonian 
or higher spin formalisms, it is still possible to define a symbol map.
Now observables are  matrices which take values in
classical phase space\cite{spo},\cite{bg}.
The Moyal bracket  of
the matrices $M_{ab}(p,x)$ and $N_{ab}(p,x)$ can be defined as
\be
\left( [M(p,x ) , N(p,x ) ]_\star \right)_{ab}  =  M_{ac}(p,x ) \star N_{cb}(p,x ) 
- N_{ac}(p,x ) \star M_{cb}(p,x ) . \lb{MM}
\ee
However, the classical limit (\ref{ocl})
 of (\ref{MM}),
in addition to the Poisson brackets of matrices, yields
a commutator of matrices  which is singular.
Generally, the observables in a block--diagonal form are taken into acoount
for getting rid of the matrix commutator.
As far as 
observables possessing a direct classical interpretation
are considered, this restriction
seems necessary for a semiclassical study\cite{bg}.
Indeed, we will relax this condition.
When  interactions  are considered, 
the nonrelativistic limit of the Dirac hamiltonian may include the spin. 
Then there will be terms depending on $\hbar $ whose
classical limit is not direct.
We would like to study the semiclassical spin dynamics. Thus, although we  deal with 
the classical phase space we let the
symbols depend on $\hbar .$ Therefore,  
 instead of the classical limit (\ref{ocl}) 
we deal with the  limit obtained 
from the Moyal bracket (\ref{MM}) by retaining the  terms 
up to  $\hbar $:  
\be
 \{M(p,x ) , N(p,x ) \}_C \equiv  
\frac{-i}{\hbar}[M,N]+
\frac{1}{2} \{ M(p,x ) , N(p,x ) \} 
-\frac{1}{2} \{ N(p,x ) , M(p,x ) \} . \lb{SCB}
\ee
We would like to emphasize that the first term is the 
commutator of matrices, it is not the quantum mechanical one.
Hence, it is not an attempt to combine the quantum commutator 
and the Poisson bracket\footnote{For 
the attempts of combining the Poisson and quantum brackets,
see \cite{Kisil} and the references given therein.}. 
Although we keep terms up to $\hbar$ order in the Moyal bracket (\ref{MM}),
remember that $M$ and $N$ can depend on $\hbar .$ In fact, 
 (\ref{SCB}) is an expansion in powers of
$\hbar$ where only the first two lowest nonvanishing terms  
are retained.

Multiplication of observables is still given by the star product (\ref{star}).
Hence the Jacobi identity which should be satisfied is given by
\bea
&\{M,\{ N,L \}_C\}_\star +  
\{N,\{ L,M \}_C\}_\star  +
\{L,\{ M,N \}_C\}_\star  = &\\
&[M, \{ N , L \}]
-[M , \{ L , N\}]
+\{M ,[N,L]\}
 -\{[N,L],M\}
-\fr{i}{\hbar}[M,[N,L]] & \\
&+ ({\rm cyclic\ permutations\ of}\  M,N,L) +{\cal O} (\hbar) =0.&
\eea
In fact one can show that
it is fulfilled up to $\hbar$ order. 
Moreover, one can observe that the Leibniz rule  defined as
\be
\lb{lru}
\{M\star N, L\}_C=\{M,L\}_C\star N+ M\star \{N,L\}_C 
\ee
is also satisfied
at the $\hbar$ order.

To define semiclassical  dynamical equations 
we propose to replace the Poisson bracket in classical dynamical equations
with the semiclassical bracket (\ref{SCB}).
Let the symbol of the
Dirac hamiltonian or its nonrelativistic
approximation be the matrix $H(p,x).$ Thus we consistently establish
\be
{\dot M } (p,x) = \{M(p,x ) , H(p,x ) \}_C ,\lb{evo}
\ee
as the time evolution of the
semiclassical observable
$M(p,x).$ It is worth to recalling that,
as it is elucidated above,  in this equation of motion
one retains the lowest two nonvanishing terms in
$\hbar .$

\section{A semiclassical constrained hamiltonian system}

When a classical system is described with a Lagrangian, the definition of canonical
momenta can yield some relations between coordinates and momenta which are
called primary constraints. 
Preserving these constraints in time may produce
some other constraints\cite{hrt}. Once all the constraints 
are derived each one can be
classified as first or  second class  due to their Poisson bracket
relations. A method of treating second
class constraints is to introduce Dirac brackets which effectively
set the constraints equal to zero. We will
consider a constrained hamiltonian system utilizing  the semiclassical
bracket (\ref{SCB}) and the dynamical equation (\ref{evo}).

Let us consider the first order lagrangian  which is a  $N\times N$ matrix:
\be
\La  = {\dot r}^\al \lp \fr{1}{2}I y_\al +\rho \A_\al (r,y) +\eta a_\al (r,y) \rp \no
-{\dot y}^\al \lp \fr{1}{2}I r_\al -\xi \B_\al (r,y) \rp -\Ha_0(r,y). \lb{oL}
\ee
Here $\al =1,\cdots, n,$ and  for  the nonrelativistic case the dot 
over the variables indicates the
derivative with respect to time $t$ and for the relativistic formalism
it is the derivative with respect to an evolution parameter  $\tau .$
$\rho , \xi ,$ and $\eta  $
are coupling constants corresponding  to the gauge fields
$\A, \B ,$ and $ a$ which are  $N\times N$ matrices. 
$I$ is the unit matrix. 
 Observe that
in (\ref{oL}) generally one cannot get rid of $\A ,\B ,$ and $a$ terms by redefining
the coordinates $r_\al$ or $y_\al .$
The definition of canonical momenta 
$$
\Pi_r^\al =\frac{\del \La}{\del \dot r_\al} ,\ \Pi_y^\al =\frac{\del \La}{\del \dot y_\al}
$$ 
leads to vanishing of the relations
\beqa
\psi^{1\al} & \equiv & (\Pi_r^\al -\fr{1}{2}y^\al )I -\rho \A^\al -\eta a^\al ,\lb{pcs1}\\
\psi^{2\al} & \equiv & (\Pi_y^\al +\fr{1}{2}r^\al )I -\xi \B^\al ,\lb{pcs2}
\eeqa
which are called primary constraints. In terms of the
canonical hamiltonian $\Ha_0 $  we need to introduce the extended hamiltonian
\be
\lb{eh}
\Ha_e =\Ha_0 +\la^\al_z\psi^z_\al ,
\ee
where $\la^\al_z$ are Lagrange multipliers and $z=1,2.$ 
To employ the semiclassical approach 
of Section 2, we identify
the canonical variables as
$p^\mu=(\Pi^\al_y,\Pi^\al_r)$ and $x_\mu =(y_\al ,r_\al ).$ 
The semiclassical brackets between the constraints can be shown to be
\beqa
\{\psi^1_\al ,\psi^1_\beta \}_C & = & \rho F_{\al \beta}
+\eta f_{\al \beta} -\fr{i\rho \eta}{\hbar} [\A_\al ,a_\beta ]
-\fr{i\rho \eta}{\hbar} [a_\al ,\A_\beta ] ,\nonumber \\
\{\psi^2_\al ,\psi^2_\beta \}_C & = & \xi G_{\al \beta},\nonumber \\
\{\psi^1_\al ,\psi^2_\beta \}_C & = & -g_{\al \beta} +
\xi\fr{\del \B_\beta}{\del r^\al}
-\rho \fr{\del \A_\al}{\del y^\beta}
-\eta \fr{\del a_\al}{\del y^\beta}  -\fr{i\xi\rho }{\hbar} [\A_\al ,\B_\beta ]
-\fr{i\xi \eta}{\hbar} [a_\al ,\B_\beta ] , \nonumber 
\eeqa
where $g^{\al\beta}$ is  the flat metric and field strengths are defined as
\beqa
f_{\al \beta} & = & 
\fr{\del a_\beta}{\del r^\al}
-\fr{\del a_\al}{\del r^\beta}
-\fr{i\eta }{\hbar} [a_\al ,a_\beta ] ,\\
F_{\al \beta} & = & 
\fr{\del \A_\beta}{\del r^\al}
-\fr{\del \A_\al}{\del r^\beta}
-\fr{i\rho }{\hbar} [\A_\al ,\A_\beta ] ,\\
G_{\al\beta}  & = &  
\fr{\del \B_\beta}{\del y^\al}
-\fr{\del \B_\al}{\del y^\beta}
-\fr{i\xi }{\hbar} [\B_\al ,\B_\beta ] .
\eeqa
Therefore,  constraints (\ref{pcs1}), (\ref{pcs2}) are second class and
the condition of preserving them in time 
\be
\lb{sco}
\{ \psi^z_\al , \Ha_e \}_C\approx 0,
\ee 
where $\approx$ indicates
that the equality is valid  up to  vanishing of  constraints,
will determine $\la^z_\al .$ 
In fact, in terms of
\be
\lb{cm}
C^{zz^\prime}_{\al \beta}=
\{\psi^z_\al , \psi^{z^\prime}_\beta\}_C	;\ 
C^{zz^{\prime \prime}}_{\al \gamma }
C_{z\prime z^{\prime\prime}}^{-1\gamma \beta}=\delta_\al^\beta \delta_{z^\prime}^z,
\ee
one can show that (\ref{sco}) leads to
\be
\lb{lm}
\la^\al_z =-\{\psi^{z^\prime}_\beta ,H_0\}_C C_{z z^\prime }^{-1 \al \beta} .
\ee 

To set effectively the second class constraints (\ref{pcs1}),(\ref{pcs2}),
equal to zero,
we introduce the semiclassical Dirac bracket 
\begin{equation}
\label{sdb}
\{M,N\}_{CD} \equiv \{M,N\}_C -\{M,\psi^z\}_C C^{-1}_{zz^\prime}\{\psi^{z^\prime},N\}_C  .
\end{equation}
Now, in dynamical equations
the semiclassical bracket of observables (\ref{SCB}) should be substituted with
the semiclassical Dirac bracket (\ref{sdb}). Observe that
the coordinates satisfy
\beqa
\{r^\al,r^\beta \}_{CD}  & = &  C_{11}^{-1\al\beta} ,\lb{rr}\\
\{y^\al,y^\beta \}_{CD}  & = &  C_{22}^{-1\al\beta} ,\lb{yy}\\
\{r^\al,y^\beta \}_{CD}  & = &  C_{12}^{-1\al\beta} .\lb{ry}
\eeqa
We omitted the unit matrix I on the left hand sides.
Obviously, $C_{12}^{-1\al\beta}=-C_{21}^{-1\al\beta}=g_{\al\beta}+\cdots ,$
thus one should consider
$r_\al$ as coordinates and $y_\al$ as the corresponding momenta.

The equation of motion of an observable $\cO (r,y)$ is given
with the extended hamiltonian as
\be
\lb{eqm}
{\dot \cO }(r,y)  =  \{\cO (r,y) ,H_e\}_C ,
\ee
in accord with the  constrained  dynamical systems. 
Plugging the solution (\ref{lm}) into (\ref{eh}) yields 
$$
\Ha_e =\Ha_0 -\{\psi^{z^\prime}_\beta ,H_0\}_C C_{z z^\prime }^{-1 \al \beta} \psi^z_\al .
$$ 
The inverse matrix elements $C_{z z^\prime }^{-1 \al \beta}$ will be obtained as a power
series in the coupling constants $\rho , \xi$ which may be identified with $\hbar .$ 
Then in the equation of motion (\ref{eqm}) we will retain the lowest two nonvanishing terms 
in $\hbar .$

\section{Spin dynamics}

Within the formulation of the previous section 
we will focus on  some different approaches of studying semiclassical
dynamics of electrons in terms of Berry gauge fields. Before considering 
specific systems let us present the general formulation where  
$a_\al=a_\al(r)$ is an  abelian gauge field  and the coupling constants are 
$\eta=e/c ,\  \xi=\hbar ,\ \rho=-\hbar . $ In our notation $e<0$ for an electron.  
The matrix 
$C^{zz^\prime}_{\al \beta}$ 
defined in (\ref{cm}) reads
\be
\lb{bCA}
C^{zz^\prime}_{\al\beta}=
\lp
\begin{array}{cc}
\frac{e}{c} f_{\al \beta } -\hbar F_{\al \beta} & -g_{\al \beta } +\hbar M_{\al \beta } \\
g_{\al \beta }-\hbar M_{ \beta \al} & \hbar G_{\al\beta}
\end{array}
\rp ,
\ee
where
\be
\lb{M}
M_{\al \beta}=\fr{\del \B_\beta}{\del r^\al}
+ \fr{\del \A_\al}{\del y^\beta}
+i [\A_\al ,\B_\beta ]  .
\ee
Obviously $M_{\al\beta}$ does not possess any symmetry or antisymmetry
with respect to the indices,
so that one should distinguish $M_{\al\beta}$ from $M_{\beta\al}.$
The inverse of (\ref{bCA}) can be calculated at the first order in $\hbar$ as
\beqa
C^{-1}_{11\al\beta} & = & \hbar G_{\al \beta}, \lb{bCAI1}\\
C^{-1}_{12\al\beta} &=& g_{\al \beta }+\hbar M_{ \beta \al} -\frac{e}{c}\hbar (Gf)_{\al 
\beta}, \\
C^{-1}_{21\al\beta} & = &-g_{\al \beta } -\hbar M_{ \al \beta } +\frac{e}{c}\hbar (fG)_{\al 
\beta}, \\
C^{-1}_{22\al\beta} &= &\frac{e}{c} f_{\al \beta } -\hbar F_{\al \beta}  +\frac{e\hbar}{c} (Mf)_{\al \beta }-\frac{e\hbar}{c}(Mf)_{\beta \al} 
-\frac{e^2\hbar}{c^2} (fGf)_{\al \beta}. \lb{bCAI2}
\eeqa
The equations of motion of the phase space variables can be obtained as
\beqa
{\dot r}^\al &  = &
\hbar \lp \fr{\del H_0}{\del r^\beta} +i [ \A_\beta  ,H_0] 
\rp G^{\al\beta}  +\lp\fr{\del H_0}{\del y^\beta} -i [\B_\beta ,H_0] \rp 
 \lp g^{\al \beta }+\hbar M^{ \beta \al} -\frac{e\hbar}{c} (Gf)^{\al\beta} \rp , \lb{gss1}
\\
{\dot y}^\al & = & 
\lp\fr{\del H_0}{\del r^\beta} +i [\A_\beta  ,H_0] 
\rp \lp-g^{\al \beta } -\hbar M^{ \al \beta } +\frac{e\hbar}{c} (fG)^{ \al\beta}\rp \no
&&
+\lp \fr{\del H_0}{\del y^\beta} -i [\B_\beta ,H_0] \rp 
\Big( \frac{e}{c} f^{ \al\beta } -\hbar F^{\al\beta  }  
+\frac{e\hbar}{c} (Mf)^{\al\beta  } 
 -\frac{e\hbar}{c}(Mf)^{\beta \al} -\frac{e^2\hbar}{c^2} (fGf)^{\al \beta } 
\Big)  ,\lb{gss2}
\eeqa
at the first order in $\hbar ,$ employing definition (\ref{eqm}).

\subsection{Phase space measure}

In \cite{xsn},\cite{b} the Berry phase  emerges because of 
keeping only lower band effects in studying the semiclassical
dynamics of Bloch electrons.
To understand this formalism
let $a_\al=a_\al(r)$ be the electromagnetic gauge field
with the coupling constant
$\eta=e/c  $  and the Berry gauge fields be
$\A_\al =0,$ and $  \B_\al =\B_\al(y)$
with $\xi=\hbar .$ 
Although, in \cite{xsn},\cite{b} only the abelian gauge
field was considered
we let $\B_\al$ be non--abelian. Hence, the matrix $C$ is given as in 
(\ref{bCA}) with 
$F_{\al \beta}=M_{\al \beta}=0,$
and at the first order in $\hbar$ the following
semiclassical Dirac brackets result
\beqa
\{r_\al ,r_\beta \}_{CD} & =& \hbar G_{\al  \beta} ,\\ 
\{r_\al ,y_\beta \}_{CD} & = &g_{\al  \beta} -\frac{e\hbar}{c} (Gf)_{\al  \beta},\\
\{y_\al ,r_\beta \}_{CD} & = &-g_{\al  \beta} +\frac{e\hbar}{c} (fG)_{\al  \beta},\\
\{y_\al ,y_\beta \}_{CD} & = & \frac{e}{c}f_{\al  \beta} -\frac{e^2\hbar}{c^2} (fGf)_{\al  \beta}.
\eeqa
Similar relations were
obtained in \cite{cnp} studying the electromagnetic
interactions of anyons. 
Now, the equations of motion of $r_\al$ and $y_\al$
can be  straightforwardly derived from 
(\ref{gss1}) and (\ref{gss2}), respectively.

Adopting the formalism of the usual
 constrained hamiltonian systems \cite{fra},\cite{sen},
the semiclassical
phase space volume element in the presence of second class constraints
is given by 
$$
\lp \prod_{\al} d \Pi^\al_r d\Pi^\al_y dy_\al dr_\al  \rp \det{^{1/2}}C 
\delta(\psi_1)\delta(\psi_2) .
$$
After eliminating $\Pi_r$ and $\Pi_y$ by employing 
constraints (\ref{pcs1}), (\ref{pcs2}) and using
(\ref{bCA}) with 
$F_{\al \beta}=M_{\al \beta}=0,$
the phase space volume element
becomes
\be
\lb{psve}
\lp \prod_{\al} dy_\al dr_\al \rp {\det}^{1/2} C= 
\lp \prod_{\al} dy_\al dr_\al \rp
\lp 1 -\fr{f_{\gamma\beta}G^{\gamma\beta}}{2} \rp .
\ee
This is the phase space volume element discussed in
\cite{xsn},\cite{b}.  Although in a
different context in \cite{xsn}  the role of second class constraints 
in defining the phase space volume element (\ref{psve}) was noted.

\subsection{Unitary transformations }

The nonrelativistic approximation
of the Dirac hamiltonian interacting with external fields
can be obtained in terms of
the Foldy--Wouthuysen unitary
transformation $U.$
In \cite{mnz} and \cite{bm2}
Foldy--Wouthuysen transformations were engaged to introduce
Berry gauge fields.
In \cite{bm2} a projector on the positive energy space $\cal P$ 
is employed to define
\be
B_i={\cal P} U\fr{\del U^\dagger}{\del y^i}.
\ee
Here $y_i$ are 
the components of the three vector $\bm y $
 and the flat metric is $g_{ij}=\delta_{ij};$ $i,j=1,2,3.$  
The Berry gauge field can be shown to be\cite{bm2}
\be
\lb{bith}
B_i=\fr{c^2\epsilon_{ijk}y_j\sigma_k}{2\lp E_p^2+mc^2E_p \rp}  \ ,
\ee 
where $E_p^2=({\bm y \cdot \bm y})c^2+m^2c^4 $
and $\s_i$ are the Pauli matrices. This is a non--abelian gauge field.
Using (\ref{bith}) in the general approach 
(\ref{rr})--(\ref{ry}),(\ref{bCAI1})--(\ref{bCAI2}) with $\A=0,\ a=0$
and  $\xi =\hbar ,$   yields
\beqa
\{y_i,y_j\}_{CD} & =& 0,\\ 
\{y_i,r_j\}_{CD} & = &-\delta_{ij} ,\\
\{r_i,r_j\}_{CD} & = &
 -i\epsilon_{ijk}\fr{c^4}{2E_p^3}\lp m\s^k +\fr{y^k({\bm y \cdot \bm \s})}{E_p+mc^2}\rp .
\eeqa
These coincide with the noncommutativity relations obtained in \cite{bm2}.

On the other hand in \cite{bb} a unitary transformation 
$U=U(r,y)$  which diagonalizes 
the  initial matrix valued hamiltonian was introduced.
Generally $U=U(r,y)$
depends on all phase space variables. Hence
one can define
the gauge fields which are non--abelian 
as $\A^G_i=-U \fr{\del U^\dagger}{\del r^i},\ 
\B^G_i=U\fr{\del U^\dagger}{\del y^i},$
with $\xi=\hbar,\rho=-\hbar.$
 Because of being
 pure gauge fields 
their field strengths vanish:
$F^G_{ij}=0,\ G^G_{ij}=0.$ However, in the adiabatic approximation
one deals with
\be
\lb{ada}
\A_i^{(ad)}\equiv {\rm diag}\lp U\fr{\del U^\dagger}{\del r^i} \rp ,\
\B_i^{(ad)}\equiv {\rm diag}\lp U\fr{\del U^\dagger}{\del y^i} \rp .
\ee
Though these are abelian gauge fields, their field strengths
$$
F^{(ad)}_{ij}  = 
\fr{\del \A^{(ad)}_j}{\del r^i}
-\fr{\del \A^{(ad)}_i}{\del r^j},\
G^{(ad)}_{ij}  = 
\fr{\del \B^{(ad)}_j}{\del y^i}
-\fr{\del \B^{(ad)}_i}{\del y^j},
$$
do no longer vanish.

The equations of motion of the phase space variables
can be read directly from (\ref{gss1}) and (\ref{gss2}) as
\bea
{\dot r}_i &  = &
\hbar  \fr{\del H_0}{\del r_j}  G^{(ad)}_{ij}  +\fr{\del H_0}{\del y_j} 
 \lp \delta_{ij}+\hbar M^{(ad)}_{ij} -\frac{e\hbar}{c} (G^{(ad)}f)_{ij} \rp , 
\\
{\dot y}_i & = & 
\fr{\del H_0}{\del r_j} \lp -\delta_{ij } -\hbar M_{ij}^{(ad)} 
+\frac{e\hbar}{c} (fG^{(ad)})_{ij}\rp \\
&&
+ \fr{\del H_0}{\del y_j} 
\lp \frac{e}{c} f_{ ij } -\hbar F^{(ad)}_{ij  }  
+\frac{e\hbar}{c} (M^{(ad)}f)_{ij} 
 -\frac{e\hbar}{c}(M^{(ad)}f)_{ij} -\frac{e^2\hbar}{c^2} (fG^{(ad)}f)_{ij } 
\rp  ,
\eea
where    $f_{ij}$ is the electromagnetic field strength, $\eta=e/c$ and
$$
M^{(ad)}_{ij}  = \fr{\del \B^{(ad)}_j}{\del r^i}
+ \fr{\del \A^{(ad)}_i}{\del y^j}.
$$
In terms of these equations of motion
one can study topological spin transport.

\subsection{Transverse spin force}

We would like to  discuss
the spin dependent dynamics  obtained in
\cite{shen} within our approach by using
the equations of motion (\ref{gss1}) and (\ref{gss2}). 
For this purpose we choose the canonical hamiltonian to be
\be
\lb{ph}
H_0= \fr{1}{2m} {\bm y}^2 + V +\mu_B \bm \s \cdot \bm B  ,
\ee
where $\mu_B=-e\hbar /2mc $ and at the first order in $\hbar$ we take
$V_{\rm eff}= V({\bm r}) +\fr{\hbar^2}{8m^2c^2}
\fr{\del^2 V({\bm r})}{\del r_i^2} \approx V. $ 
$B_i=\fr{1}{2}\e_{ijk}f^{jk}$ is the external magnetic field
and $\s_i$ are the Pauli matrices.
In accord with \cite{shen} we  let 
$\B =0 $  and  the other Berry connection be
\be
\lb{AGF}
\A_i=\fr{\e_{ijk}\s_j}{4m c^2} \fr{\del V}{\del r_k} . 
\ee 
Moreover, we set $\eta =e/c,$ and $ \rho =-\hbar .$
We would like to emphasize that 
$r_i$ and $y_i$ are the coordinates and momenta in the restricted phase
space. When we plug the gauge field (\ref{AGF}) and
the canonical hamiltonian (\ref{ph}) into the equations of motion
(\ref{gss1}),(\ref{gss2}) we obtain
\beqa
\dot r_i & = &\fr{\del H_0}{\del y_i}= \fr{ y_i}{m} ,\lb{xd} \\
\dot y_i & = & -\fr{\del H_0}{\del r_i}-i[\A_i,H_0]
+\fr{\del H_0}{\del y_j}\lp \fr{e}{c} f_{ij} -\hbar F_{ij}\rp . \lb{yd}
\eeqa
The force can directly be read from (\ref{yd}) in terms of the velocity 
${\bm v} \equiv \dot {\bm r}$ given by (\ref{xd}) as
\beqa
& \F_i  =\dot y_i =m \ddot{r}_i = &
-\fr{\del }{\del r_i}\lp V +\mu_B
{\bm \s \cdot \bm B} \rp 
+ \frac{e}{c}\e_{ijk} v_jB_k   
+\frac{\hbar}{4mc^2} \lp \e_{ijk}\s^jv_l\frac{\del^2 V}{\del r_l \del r_k}
 +\e_{jkl}\s^j v^k\frac{\del^2 V}{\del r_l \del r_i}\rp   \no
&&
  +\fr{\mu_B}{2mc^2}\lp\s_i B_l\fr{\del V}{\del r_l}
 -B_i \s_l \fr{\del V}{\del r_l} \rp  
+\fr{\hbar}{8m^2c^4}\e_{ijk} \s_l
\fr{\del V}{\del r_l} v_j\fr{\del V}{\del r_k} . \lb{for}
\eeqa
Indeed, this is the  force  obtained in \cite{shen}. The last term is
the transverse spin force on the spin current quadratic in the electric field. 

Equations of motion following from the
nonrelativistic approximation of the Dirac hamiltonian
can be derived in electrodynamics employing the 
Thomas precession\cite{jac} without referring to the Dirac hamiltonian. 
The relation between the 
nonrelativistic limit and the
Thomas precession was clarified in \cite{mat} by showing that the latter
should be considered as a Berry phase when 
the external electric potential is smooth. 
The related gauge field can be obtained in the
nonrelativistic limit from (\ref{bith}).
Hence, to obtain the force acting on an electron in the external electric and
magnetic fields it should be possible to consider either the gauge field
$\A$ given in (\ref{AGF}) or the gauge field $\B$ obtained from (\ref{bith}) in 
the nonrelativistic limit: Let  $\A=0$ and  deal with the electrodynamic gauge field $a_i 
(r),\ \eta=e/c$
and the nonrelativistic limit of the non--abelian gauge field (\ref{bith})  
\be
\lb{nrb}
\B_i=\fr{1}{4m^2c^2}\e_{ijk}y^j\s^k.
\ee
Field strength of this gauge field can be calculated to be
\be
 G_{ij}=\fr{-1}{2m^2c^2}\epsilon_{ijk} \s^k 
 +\frac{1}{8m^4c^4} \epsilon_{ijk} y_k (\bm \s \cdot \bm y ),
\ee
where we used $\xi=\hbar.$

By ignoring the $fG$ terms in (\ref{gss1}) and (\ref{gss2}) the
equations of motion  are
\beqa
\dot y_i & = & -\frac{\del (V +\mu_B \bm \s \cdot \bm B)}{\del r^i} 
+\frac{e}{mc}\epsilon_{ijk} y_j B_k , \\
\dot r_i&=& \frac{y_i}{m}+\hbar G_{ij}\frac{\del V}{\del r_j} 
+\frac{\m_B}{2m^2c^2}\Big(  B_i (\bm y \cdot \bm \s )  
- \s_i (\bm B \cdot \bm y )\Big) .
\eeqa

Now, by keeping the terms linear in the velocity $v_i$ 
one can show that
\beqa
m \ddot{r}_i = m\{\dot r_i , H_e\}_C &=& -\fr{\del }{\del r_i}\lp V 
+\mu_B \bm \s \cdot \bm B \rp  
+ \frac{\hbar}{2mc^2}\e_{ijk}\s^j v_l
\frac{\del^2 V}{\del r_l \del r_k} 
+\frac{e}{c}\e_{ijk} v_jB_k \no
&& +\fr{\mu_B}{2mc^2}\lp\s_i B_l\fr{\del V}{\del r_l}
 -B_i \s_l \fr{\del V}{\del r_l} \rp 
 +\fr{\hbar}{8m^2c^4}\e_{ijk} \s_l
\fr{\del V}{\del r_l} v_j\fr{\del V}{\del r_k} .   \lb{leq}
\eeqa 
Up to some $\del^2 V/\del r_i\del r_j$ terms this coincides
with (\ref{for}). In fact, the latter approach is valid for
potentials changing slowly. 
However, neglecting the $fG$ terms in (\ref{gss1}),(\ref{gss2})
is not justified, due to the fact that they may give contributions
of the   $\mu_B /mc^2$ order to the force. Indeed, retaining
the $fG$ terms in (\ref{gss1}),(\ref{gss2}) and
using $\mu_B=-e\hbar /2mc ,$
the equations of motion of the $\hbar$ order are
\beqa
\dot y_i & = & -\frac{\del (V +\mu_B\bm \s \cdot \bm B)}{\del r^i} 
+\frac{e}{mc}\epsilon_{ijk} y_j B_k 
-\frac{\mu_B}{mc^2}\left( \frac{\del V}{\del r^i} \s_j  B^j 
-\s_iB_j \frac{\del V}{\del r_j} \right) ,\\
\dot r_i&=& \frac{y_i}{m}+\hbar G_{ij}\frac{\del V}{\del r_j} 
-\frac{\m_B}{2m^2c^2}\Big(  B_i (\bm y \cdot \bm \s ) 
+\s_i(\bm y \cdot \bm B) 
- 2y_i (\bm B \cdot \bm \s )\Big) .
\eeqa
Hence, the force linear in velocity becomes
\beqa
m \ddot{r}_i& = &-\fr{\del }{\del r_i}\lp V 
+\mu_B \bm \s \cdot \bm B \rp  
+ \frac{\hbar}{2mc^2}\e_{ijk}\s^j v_l
\frac{\del^2 V}{\del r_l \del r_k} 
+\frac{e}{c}\e_{ijk} v_jB_k \no
&& +\fr{\mu_B}{2mc^2}\lp3\s_i B_l\fr{\del V}{\del r_l}
 +B_i \s_l \fr{\del V}{\del r_l} -4\fr{\del V}{\del r_i}(B_l\s_l)\rp 
 +\fr{\hbar}{8m^2c^4}\e_{ijk} \s_l
\fr{\del V}{\del r_l} v_j\fr{\del V}{\del r_k} .   \lb{leq2}
\eeqa 
The last term which is 
the transverse spin force
on the spin current results to be the same\footnote{In \cite{Bliokh}
it was claimed that this method 
leads to a transverse force in conflict with \cite{shen}. }.
However, the terms which depend on
 both the electric and magnetic fields are in dispute with
(\ref{for}). This discrepancy between the two   nonrelativistic
approximation schemes can be  settled  by experiments.

\subsection{The Spin Hall Effect}

Electrons constrained to move on a plane in the presence of a uniform  external
magnetic field perpendicular to the plane  deviate and produce an electric
field which is perpendicular  to both the initial direction
of the current and the magnetic field. This is the Hall effect which 
 manifests itself as the Hall conductivity. We would like to
present a derivation of the Hall conductivity
which will inspire  a  simple formulation
of the intrinsic spin Hall effect utilizing our semiclassical approach.
To this aim let us deal with the  hamiltonian  
\be
\lb{hal0}
H_0= \fr{1}{2m} {\bm y}^2 + V (r_1,r_2) ,
\ee
where the scalar potential is given in terms of the uniform electric 
field components $E_i$ as
\be
V (r_1,r_2)=-eE_1r_1-eE_2r_2 .
\ee
In order to constrain the electron to move on $r_1r_2$--plane we set $y_3=0.$
We consider the vanishing  Berry gauge fields  $\A =0,\B =0$
and  let there be a uniform magnetic field in $r_3$ direction:
\be
f_{12}=B.
\ee
The related coupling constant is $\eta =e/c .$
The equations of motion following from (\ref{gss1}) and (\ref{gss2}) are
\beqa
\dot r_i & = & \fr{ y_i}{m} ,\lb{hal11} \\
\dot y_1 & = & eE_1 + \fr{eB}{mc}  y_2 , \lb{hal12}\\
\dot y_2 & = & eE_2 -\fr{eB}{mc}  y_1, \lb{hal13}
\eeqa
The force acting on  electron  can be read from (\ref{hal11})--(\ref{hal13}),
in terms of the velocity 
${\bm v} \equiv \dot {\bm r},$ as
\beqa
\F_1 = m \ddot{r}_1 =\dot y_1 = e E_1+\fr{eB}{c}v_2, \lb{half}\\
\F_2 = m \ddot{r}_2 =\dot y_2 = e E_2-\fr{eB}{c}v_1. \lb{half1}
\eeqa
Till now we have considered single particle dynamics. 
To connect it to a system of
electrons let us introduce  the density of electrons $\kappa$. Thus,
the electric current  is defined by
\be
\lb{cure}
{\bm j}=e\kappa {\bm v}.
\ee
We demand that the net force acting on electrons vanish $ \F_i =0,$ so that 
the electrons move without deflection (see e.g. \cite{gir}). We
can solve this condition for the velocity and plug it into (\ref{cure}),
which yields  the electric current
\be
\left(
\begin{array}{c}
j_1\\
j_2	
\end{array} \right)
=\left(
\begin{array}{cc}
0 & -\sigma_H\\
\sigma_H & 0	
\end{array}
\right)
\left(
\begin{array}{c}
E_1\\
E_2	
\end{array}
\right)
\ee
where
$$
\sigma_H =-\frac{ec\kappa}{B}
$$
is the Hall conductivity.

The intrinsic spin Hall effect is envisaged in \cite{sin} in terms of the  Rashba
spin--orbit coupling\cite{ras}. 
By generalizing the Hall effect formulation we can introduce
a simple method of acquiring the spin 
Hall effect conductivity employing the Rashba spin--orbit coupling. 
The  hamiltonian is still given by
(\ref{hal0}) with $y_3=0.$ However, there is no magnetic field: $a_i=0.$ 
To consider the linear Rashba spin--orbit coupling we set $\B =0 $  and define  
\be
\lb{ars}
\A_i=\e_{ijk}\s_j  e^z_k .  
\ee 
Here $\bm e^z$ is the unit vector in the third direction
$e^z_k =\delta_{k3}$ and $\s_i$ are  the Pauli  matrices.
Moreover, in the original formulation (\ref{oL}) we should take
 $ \rho =-\al m/ \hbar ,$ where $\al$ is the Rashba coupling constant\cite{ras}.

The related field strength can be calculated as
\be
\lb{fra}
F_{ij}=-\frac{i\rho}{\hbar}[\A_i,\A_j]=\frac{2\rho}{\hbar} 
\sigma_3 \e_{ijk} e^z_k .
\ee

The equations of motion of the canonical variables are
\beqa
\dot r_i & = & \fr{ y_i}{m} , \\
\dot y_i & = & -\fr{\del V}{\del r_i} +\fr{\rho}{m} F_{ij} y_j .
\eeqa
Hence, the force acting on the particle is 
\be
\lb{raf1}
\F_i=m\ddot r_i  =eE_i  + \frac{2\rho^2}{\hbar} \sigma_3\e_{ijk} e^z_k v_j.
\ee
Imitating the formulation of the Hall effect we 
set  $\F_i =0,$ in order to have a motion without deflection.  
This condition is solved for  the velocity as
\beqa
&v_1^\uparrow  = \frac{e\hbar}{2\rho^2}E_2, \lb{v1u} 
v_1^\downarrow  =  -\frac{e\hbar}{2\rho^2}E_2, &\\
&v_2^\uparrow  = -\frac{e\hbar}{2\rho^2}E_1,  
v_2^\downarrow  =  \frac{e\hbar}{2\rho^2}E_1. & \lb{v2d}
\eeqa
The arrows 
$\uparrow$ and $\downarrow$ correspond, respectively, to the positive 
and negative eigenvalue of $\s_3.$ It is natural to define 
the spin current   as
\be
\lb{cur}
{\bm j}^z=\frac{\hbar}{2}
\left(n^\uparrow {\bm v}^\uparrow - n^\downarrow {\bm v}^\downarrow \right),
\ee
where $n^\uparrow$ and $n^\downarrow$ denote the concentrations of states with spins 
along the $e^z$ and $-e^z$ directions.  Employing (\ref{v1u}),(\ref{v2d}) in 
(\ref{cur}) yields
\be
\lb{cur1}
{\bm j}^z =\s_{SH}   {\bm e^z} \times {\bm E},
\ee  
where
\be
\lb{shc}
\s_{SH}=\frac{-e\hbar^2}{4\rho^2}\left(n^\uparrow +n^\downarrow\right)
\equiv \frac{-e\hbar^4}{4\al^2m^2}n
\ee
is the spin Hall conductivity. 
In this simplistic approach
the total concentration of states $n=\left(n^\uparrow +n^\downarrow\right)$
is an input which should be given by other means. Although its calculation is 
beyond the scope of this work, for having an insight let
\be
\lb{n}
n=sn_{2D}^*
\ee
where $s$ is a constant and $n_{2D}^*$ is the concentration of states occupying the 
lower energy state of the Rashba hamiltonian\cite{ras}:
$$
n_{2D}^*= \frac{\al^2 m^2}{\pi \hbar^4}.
$$
Using (\ref{n}) in (\ref{shc}) leads to the spin Hall conductivity
\be
\s_{SH}= -\frac{es}{4\pi}.
\ee
This agrees with the universal behavior obtained in \cite{sin} for  $s=1/2.$
However, when the vertex corrections are taken into account 
it is known that
this universal behavior does not survive\cite{vsh},
as far as the linear Rashba coupling is considered.
The vertex corrections were calculated employing Green functions
within the Born approximation.
Hence, it is not clear how one can incorporate the vertex corrections into our semiclassical scheme.
To cure the defects of the linear theory, it would be useful
to study the  Rashba couplings which are higher orders in momenta (see \cite{erh} and 
the references therein). 
Although we will not discuss it here,
our semiclassical approach can be used to investigate higher order generalizations
of the Rashba spin--orbit coupling. 

\section{Discussions}

Semiclassical limit designated as the bracket (\ref{SCB}) can be utilized
to study diverse dynamical problems where spin degrees of freedom are not ignored. 
Hence, instead of dealing with wave packets one can consider single particle interpretation
of semiclassical dynamics of spin dependent systems. 

It can be shown that the constrained hamiltonian system which we presented here 
is suitable to investigate properties of some topological quantum phases.
Moreover, as we will present in a future work it constitutes 
a new gauge invariant method of studying dynamical systems
in  noncommutative spaces. 

Any model concerning spin dynamics utilizing
Berry gauge fields which give rise to noncommutativity
of coordinates and/or momenta  can be studied
in terms of the semiclassical approach presented here. We focused on
some recent formalisms 
where some of persisting discussions can be clarified.
The results which we derived are valid up to  the first order in $\hbar .$
However, in this formulation keeping the track of the higher orders is possible.
When higher order $\hbar$ corrections are considered there may be some
different sources: Gauge fields which we consider may depend on higher
$\hbar ,$ the   limit of the Moyal bracket will have another
term and inversion of the matrix $C_{\al \beta}^{zz^\prime}$ 
may lead to some higher $\hbar$ terms.

Obviously, the formalism of the spin hall conductivity which we reported here
should be elaborated. Nevertheless, 
due to its resemblance with the Hall effect
and simplicity, it may be profitable to predict
some basic properties of the spin Hall effect.

\vspace{1cm}

\begin{center}
{\bf Acknowledgment}
\end{center}
I would like to thank M. Elbistan for fruitful discussions.

\end{document}